\documentstyle{mn}

\newif\ifAMStwofonts

\ifoldfss
  \ifCUPmtlplainloaded \else
    \NewTextAlphabet{textbfit} {cmbxti10} {}
    \NewTextAlphabet{textbfss} {cmssbx10} {}
    \NewMathAlphabet{mathbfit} {cmbxti10} {} 
    \NewMathAlphabet{mathbfss} {cmssbx10} {} 
  \fi
  \ifAMStwofonts
    \ifCUPmtlplainloaded \else
      \NewSymbolFont{upmath} {eurm10}
      \NewSymbolFont{AMSa} {msam10}
      \NewMathSymbol{\upi}     {0}{upmath}{19}
      \NewMathSymbol{\umu}     {0}{upmath}{16}
      \NewMathSymbol{\upartial}{0}{upmath}{40}
      \NewMathSymbol{\leqslant}{3}{AMSa}{36}
      \NewMathSymbol{\geqslant}{3}{AMSa}{3E}

    \fi
  \fi
\fi 

\ifnfssone
  \newmathalphabet{\mathit}
  \addtoversion{normal}{\mathit}{cmr}{m}{it}
  \addtoversion{bold}{\mathit}{cmr}{bx}{it}
  \newmathalphabet{\mathbfit} 
  \addtoversion{normal}{\mathbfit}{cmr}{bx}{it}
  \addtoversion{bold}{\mathbfit}{cmr}{bx}{it}
  \newmathalphabet{\mathbfss} 
  \addtoversion{normal}{\mathbfss}{cmss}{bx}{n}
  \addtoversion{bold}{\mathbfss}{cmss}{bx}{n}
  \ifAMStwofonts
    \ifCUPmtlplainloaded \else
      \UseAMStwoboldmath
      \makeatletter
      \new@mathgroup\upmath@group
      \define@mathgroup\mv@normal\upmath@group{eur}{m}{n}
      \define@mathgroup\mv@bold\upmath@group{eur}{b}{n}
      \edef\UPM{\hexnumber\upmath@group}
      \new@mathgroup\amsa@group
      \define@mathgroup\mv@normal\amsa@group{msa}{m}{n}
      \define@mathgroup\mv@bold\amsa@group{msa}{m}{n}
      \edef\AMSa{\hexnumber\amsa@group}
      \makeatother
      \mathchardef\upi="0\UPM19
      \mathchardef\umu="0\UPM16
      \mathchardef\upartial="0\UPM40
      \mathchardef\leqslant="3\AMSa36
      \mathchardef\geqslant="3\AMSa3E
    \fi
  \fi
\fi 

\ifnfsstwo
  \DeclareMathAlphabet{\mathbfit}{OT1}{cmr}{bx}{it}
  \SetMathAlphabet\mathbfit{bold}{OT1}{cmr}{bx}{it}
  \DeclareMathAlphabet{\mathbfss}{OT1}{cmss}{bx}{n}
  \SetMathAlphabet\mathbfss{bold}{OT1}{cmss}{bx}{n}
  \ifAMStwofonts
    \ifCUPmtlplainloaded \else
      \DeclareSymbolFont{UPM}{U}{eur}{m}{n}
      \SetSymbolFont{UPM}{bold}{U}{eur}{b}{n}
      \DeclareSymbolFont{AMSa}{U}{msa}{m}{n}
      \DeclareMathSymbol{\upi}{0}{UPM}{"19}
      \DeclareMathSymbol{\umu}{0}{UPM}{"16}
      \DeclareMathSymbol{\upartial}{0}{UPM}{"40}
      \DeclareMathSymbol{\leqslant}{3}{AMSa}{"36}
      \DeclareMathSymbol{\geqslant}{3}{AMSa}{"3E}
    \fi
  \fi
\fi 

\ifCUPmtlplainloaded \else
  \ifAMStwofonts \else 
    \def\upi{\pi}
    \def\umu{\mu}
    \def\upartial{\partial}
  \fi
\fi

\title{Inverse Compton scattering model for gamma-ray production 
in MeV blazars}
\author[W. Bednarek]
       {W. Bednarek\\
Department of Experimental Physics, University of \L\'od\'z, 90-236 \L\'od\'z, 
ul. Pomorska 149/153, Poland.  
               }
\date{Accepted 199 Received 199; in original form 1997}

\pubyear{199}

\begin{document}

\maketitle

\label{firstpage}

\begin{abstract}

It is proposed that the spectra of 
so called 'MeV blazars' can be explained in terms of previously 
developed models of the external comptonization of accretion  disk radiation 
provided that the 
structure of the inner and outer parts of the accretion disk is different. 
The electron acceleration is saturated by the inverse Compton losses in the
inner geometrically thick disk and the outer geometrically thin disk at 
different maximum energies which causes the appearance of two spectral
components, one strongly peaked in the MeV energy range and the second one 
of a power law type extending through the GeV energy range.
The spectra, computed in terms of such a simple geometrical model, are in a good 
agreement with observations of the MeV blazar PKS0208-512. They are consistent 
with the transient appearance of a strong MeV peak, the power law 
spectrum in the EGRET energy range, and a possible cut-off at high energies.

\end{abstract} 


\section{Introduction}

The COMPTEL telescope on the board of Compton GRO has detected a few blazars 
which emitted power is concentrated in the MeV energy range 
(Blom et al.~1995, Bloemen et al.~1995). Two spectral components can be 
clearly identified in such type of sources (e.g. PKS0208-512, Kanbach 1996, 
Blom et al.~1996). It has been proposed that such spectra can originate in a 
dense $e^{\pm}$ relativistic blob moving in a jet. In such  model
the MeV peak is caused by the blueshifted $e^{\pm}$ annihilation line
(Marcowith, Henri \& Pelletier~1995, Roland \& Hermsen~ 1995, B\"ottcher 
\& Schlickeiser~1996, Skibo, Dermer \& Schlickeiser~1997). However 
observations of 3C 273 by the OSSE telescope show that in fact the peak can 
be sometimes present at $\sim
0.3$ MeV, which would rather require the redshifted $e^{\pm}$ annihilation 
line (McNaron-Brawn et al. 1997). Recently it has been suggested that these 
two emission components may be created 
by two different populations of electrons present in a jet, which scatter
the external quasi-isotropic soft radiation (Sikora \& Madejski~1996). 

It has been already noted that the spectra of MeV blazars could be explained
by the emission of relativistic electrons which comptonize the thin disk radiation
(Bednarek, Kirk, Mastichiadis 1996a,b (BKM96a,b)). However, recently observed 
very flat transient spectrum in the GeV energy range with a cut-off
at a few GeV (Kanbach 1996) is difficult to explain with the assumption on the 
thin disk geometry.
In this paper we show that different stages of emission of MeV blazars can be 
explained in terms of previously developed model provided that the 
structure of the inner and outer parts of the accretion disk is different. 
The electron acceleration caused, either by a shock propagating along the jet
or by electric fields induced in the reconnection regions,
can be saturated by inverse Compton losses. However, if the geometry of 
the accretion disk, surrounding central black hole, changes from the thin one
(the outer disk part) to the thick one (the inner disk part) and the inner 
parts of the disk are relatively hotter, then the acceleration of 
electrons saturates at different energies during jet propagation through
these radiation fields. It results in  an appearance of two spectral 
components, the first one strongly peaked in the MeV energy range and the 
second one, of a power low type, extending through the GeV energy range.

\section{Gamma-rays from scattering of photons from thick and thin 
disk}

We consider the standard external Compton scattering model for the 
$\gamma$-ray production in active galaxies, in which  
electrons accelerated in the jet lose energy on inverse Compton scattering 
(ICS) process by scattering the accretion disk radiation. 
In general, the accretion disk may be geometrically thin (as 
considered by e.g. Dermer, Schlickeiser \& Mastichiadis~1992, Dermer \& 
Schlickeiser~1993, Coppi, Kartje \& K\"onigl~1993, BKM96a,b), or it may be 
geometrically thick (Melia \& K\"onigl~1989, Bednarek~1993, 
Bednarek \& Kirk 1995 (BK95)). The electrons in turn may be accelerated
by a relativistic shock moving in the jet (e.g. Dermer \& Schlickeiser~1993,
Sikora, Begelman \& Rees~1994, Blandford \& Levinson~1995), or they move 
almost rectilinearly along the jet axis (Melia \& K\"onigl~1989, 
Coppi, Kartje \& K\"onigl~1993, BK95, BKM96). In this paper we 
consider the simpler case, assuming that the electrons are accelerated 
rectilinearly in the jet in a relatively small localised regions, either in 
electric fields
induced in magnetic reconnection or by drifting along the surfaces of 
highly oblique shocks. Such shocks may be formed when the jet plasma meets  
obstacles e.g., winds of massive stars moving through the jet 
(Bednarek \& Protheroe~1997) or debris of close supernova explosion 
(Bednarek 1997). These acceleration regions are located along the axis of 
the jet in our model.

In the present paper we apply the model for the disk geometry considered 
previously by Melia \& Konigl~(1989). Therefore, we assume that the outer disk, 
at distances greater than $r_{mid}$ from the central black hole, is 
geometrically thin and is characterised by the Shakura \& Sunyaev~(1973) 
temperature profile. The inner geometrically thick disk (disk funnel), extends
between $r_{\rm in}$ and $r_{\rm mid}$. The funnel has an opening angle 
$\theta$. For the outer disk we assume the Shakura \& Sunyaev temperature 
profile of the form
\begin{eqnarray}
T_{\rm out}(r) = T_{\rm mid} (r/r_{\rm mid})^{-0.75},  
\end{eqnarray}
\noindent
where $T_{\rm mid}$ is the surface temperature of the disk at the distance 
$r_{\rm mid}$ from the disk centre. For the temperature profile of the inner, 
thick disk, we assume that it has a similar form as the outer disk 
but with different exponent
\begin{eqnarray}
T_{fun}(r) = T_{mid} (r/r_{mid})^{-\delta}.  
\end{eqnarray}
\noindent
We discuss the cases with $\delta > 0.75$, which means that the inner disk 
is hotter than expected in Shakura \& Sunyaev model. It is worth to note that 
this is consistent with the observations of the soft X-ray excess in many 
active galaxies, between them 3 C273 which is also detected in $\gamma$-rays
(Lichti et al. 1995). Hence the disk radiation field is defined by four 
parameters: $T_{mid}$, $r_{mid}$, $\delta$, and $\theta$. The
distribution of the electric field along the jet is defined following BKM96b,
\begin{eqnarray}
E(z) = E_0 (z/z_0)^{-\alpha},
\end{eqnarray}
\noindent
where $E_0$ is the electric field strength at the distance $z_0$ along the 
jet, and the exponent $\alpha$ is a parameter. 

   \begin{figure}
      \vspace{7.cm}
\includegraphics{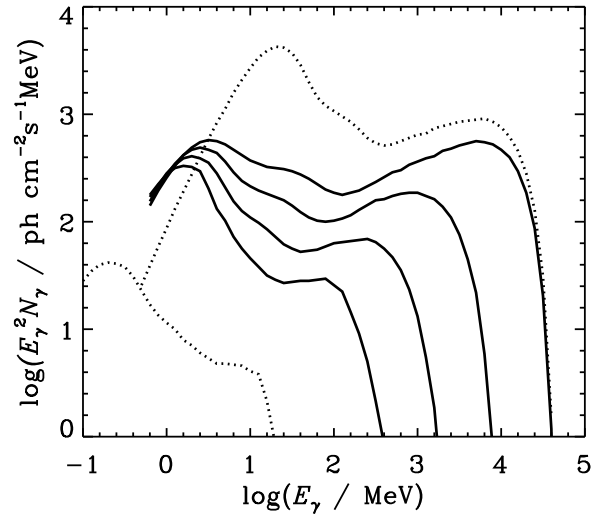}
      \caption[]{The ICS $\gamma$-ray spectra produced by electrons which 
scatter the radiation of a thick and thin accretion disk. The parameters of 
the disk are following: $T_{\rm mid} = 5\times 10^4$ K, 
$r_{\rm mid} = 2 r{\rm in}$, $r_{\rm in} = 10^{14}$ cm, $\delta = 2$, and 
$\theta = 80^0$. The spectra marked by the full curves are 
obtained for $E_0 = 0.08$ V cm$^{-1}$, $l_{\rm rec} = 0.005$, and different 
values of $\alpha = 0.67, 1., 1.33, 1.67$ (counting from the top). 
The dotted curves show the spectra for the same value of $l_{\rm rec}$ but: 
$E_0 = 8\times 10^{-3}$ V cm$^{-1}$ (bottom dotted curves), 
$E_0 = 0.8$ V cm$^{-1}$ (upper dotted curve), and $\alpha = 1.33$. }
\label{fig1}
    \end{figure}

The Lorentz factor of electrons in the jet, $\gamma_{eq}$, is determined by 
the balance
between the energy gains from the rectilinear accelerator and the energy losses
on ICS in the disk radiation as defined in BK95, and BKM96a,b, 
\begin{eqnarray}
P_{ICS}(\gamma_{eq}(z)) = c e E(z),
\end{eqnarray}
\noindent
where $P_{\rm ICS}$ are the ICS losses of electrons with Lorentz factor 
$\gamma_{\rm eq}$ at a distance $z$, $c$ is the velocity of light, and $e$ is 
the elementary electric 
charge. Bednarek, Kirk \& Mastichiadis~(BKM96a,b) has shown that the jet 
can be separated in two regions. In the inner part, called the ``radiation 
dominated jet'', electron energy losses are strong enough that they can 
reach the equilibrium Lorentz factor determined by the balance between energy 
gains and losses. In the outer part, called the ``particle dominated jet'', 
the ICS losses are too small and electrons are injected into
the jet with the Lorentz factors corresponding to the maximum potential drop
in the acceleration regions. 
These two regions are separated at $z_{max}$, measured along the jet,  which 
is given by the condition (presented in BKM96a,b),
\begin{eqnarray}
m_ec^2 \gamma_{eq}(z_{max}) = e l_{coh}E(z_{max}),
\end{eqnarray}
\noindent
where $l_{coh} = \beta z_{max}$ is the coherence length for the 
acceleration of electrons (it can be scaled by the distance from the black 
hole with the scaling factor $\beta << 1$), $E(z_{max})$ is the electric field 
in the acceleration region (given by Eq.~(3)), 
and $m_e$ is the electron rest mass.
In the radiation dominated jet, the energy of electrons is directly transferred
to the $\gamma$-ray photons in ICS process. In the particle dominated jet
the electrons are likely to be isotropized by the random component of the 
magnetic field after emerging from the acceleration regions. These electrons 
lose energy mainly on synchrotron process
contributing to the lower energy part of the blazar spectrum.

   \begin{figure*}
      \vspace{7.2cm}
\includegraphics{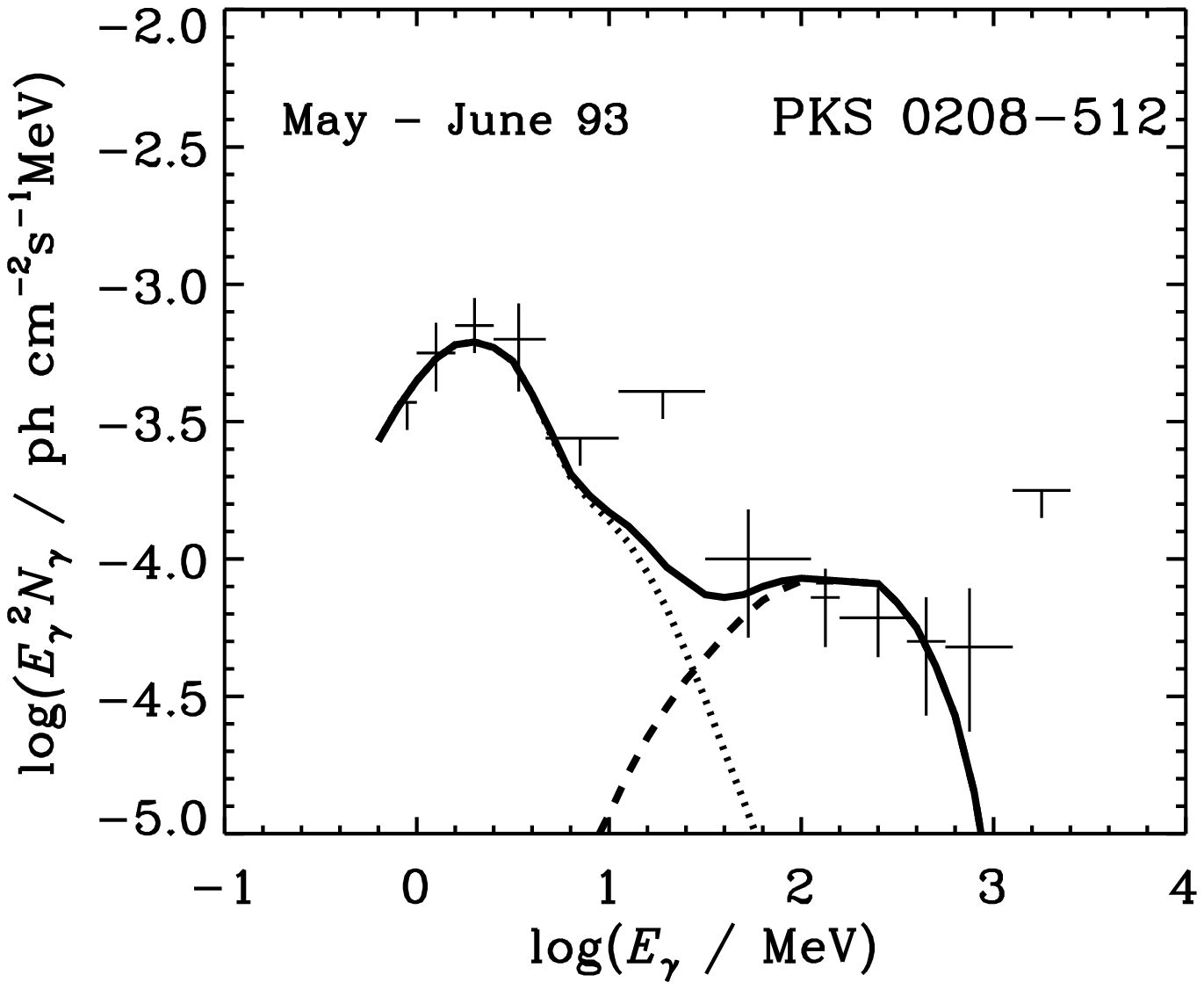}
\includegraphics{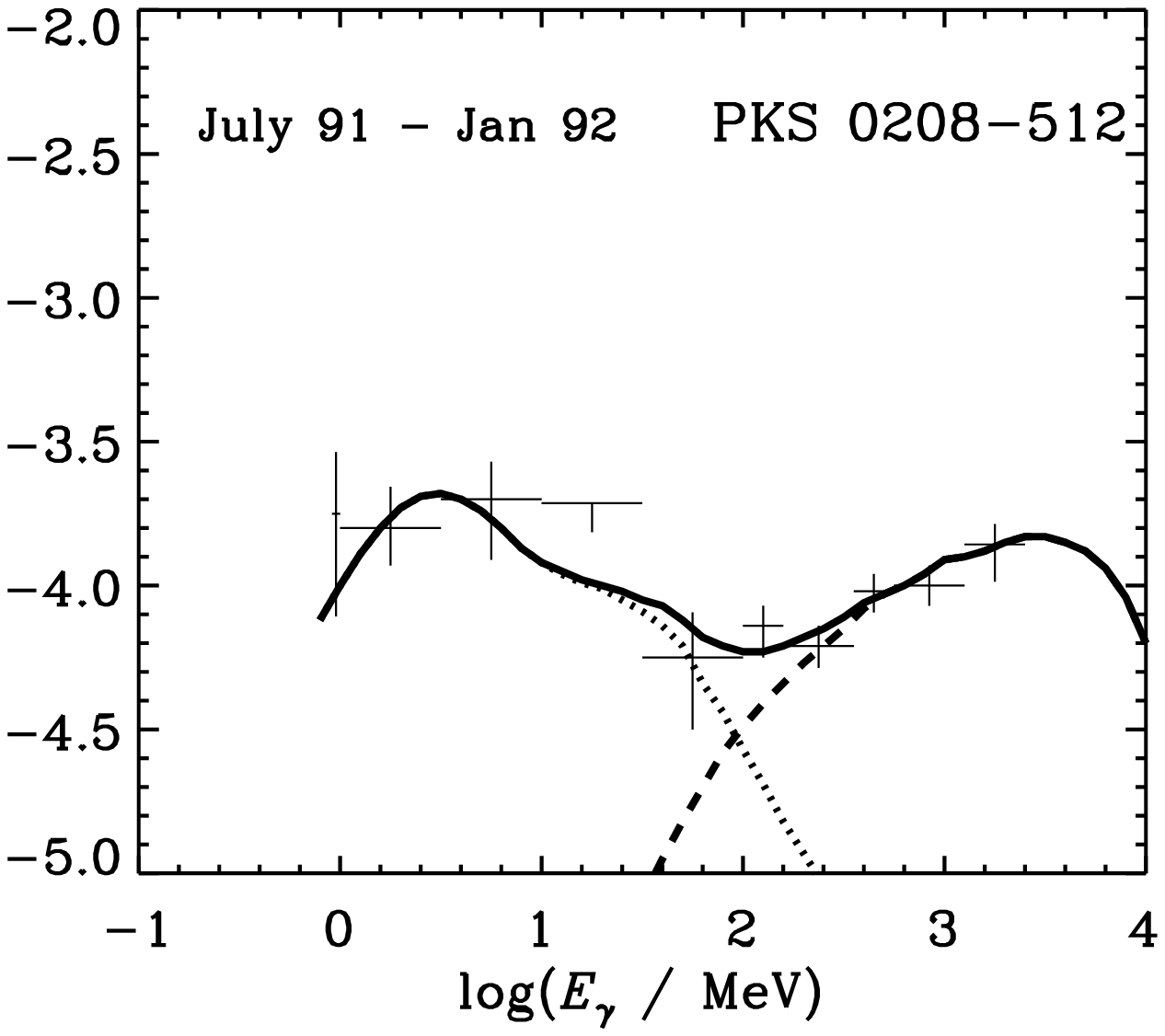}
      \caption[]{The spectra of $\gamma$-rays from ICS, produced 
by electrons propagating through radiation of the inner geometrically thick 
disk and the outer geometrically thin disk, are compared 
with the observations of MeV blazar PKS 0208-512 (Blom et al.~1996) during 
May - June 93 (on the left) and July 91 - Jan 92 (on the right). The 
differences between these spectra are only caused by the change in the 
distribution of the electric field profile along the jet defined by 
the parameter $\alpha$,
which is equal to $\alpha = 1.4$ for May - June 93 observations, and 
$\alpha = 0.67$ for July 91 - Jan 92 observations. For the other parameters 
of the fit see text.}
\label{fig2}
    \end{figure*}

In general, electrons accelerated in the electric fields may scatter disk 
radiation in the Thomson or the Klein-Nishina case.
Here we concentrate on the Thomson regime. If the scattering occurs in a narrow 
disk funnel, the Lorentz factor of electrons has to be limited to the values 
determined by the condition $\gamma_{eq} (3k_B T/m_ec^2) < 1$, where $k_B$ is 
the Boltzmann constant and $T$ is given by Eq~2. If the scattering occurs 
above the surface of a thin disk, the condition derived in BKM96a (see Eq.~(5) 
in that paper) has to be fulfilled.

The spectra of $\gamma$-ray photons produced in the radiation dominated jet 
are obtained by using the following formula
\begin{eqnarray}
{{dN}\over{dE_{\gamma} d\Omega dt}} = {{r_{in}}\over{c}} 
\int_{z_{0}}^{z_{max}} {{dN_e}\over{d\Omega dt}} {{dN(\gamma_{eq}(z))}
\over{dE_{\gamma}dt}} dz,
\end{eqnarray}
\noindent
where $dN_e/d\Omega dt$ is the injection rate of electrons (per sr$^{-1}$ 
s$^{-1}$), $dN(\gamma_{eq}(z))/dE_{\gamma}dt$ 
is the ICS $\gamma$-ray spectrum produced by an electron, with Lorentz 
factor $\gamma_{eq}(z)$, located at the distance $z$ along the jet. 
This ICS spectrum is obtained by integration 
of Eq. (2.33) in Blumenthal \& Gould~(1970) over the disk radiation field. 
$z_{0}$ is taken to be equal to $r_{in}$, and $z_{max}$ is determined by
Eq.~(5).

In Fig.~1 we show an example of the $\gamma$-ray spectra calculated in terms 
of this model. In order to obtain these spectra we fix the disk parameters and 
investigate 
their dependence on the jet parameters. Note that the spectra show two clear
components, corresponding to propagation of the jet through the radiation
of the inner, geometrically  thick disk (lower energy part of the spectrum) 
and propagation through the radiation of the outer, geometrically thin disk
(higher energy part of the spectrum). The parameter $\alpha$, which describes 
the profile of the electric field in the jet, determines the shape of the
spectrum (see full curves in Fig.~1). The location of the maximum in the 
power spectrum is determined by the strength of the electric field at the 
base of the jet (see dotted curves in Fig.~1). 

\section{Comparison with the spectrum of PKS0208-512}

PKS0208-512, at the distance corresponding to the redshift $z = 1.003$, 
has been observed by the Compton GRO a few times (e.g. Blom et al.~1996)
showing complex and variable spectrum. In some observations very strong peak 
was detected in $1\div 3$ MeV energy range by the COMPTEL detector. 
The power law spectral component extending above $\sim 30$ MeV, with the 
cut-off at a few GeV, has been also detected by EGRET detector (Kanbach~1996). 
The spectrum is variable 
with the spectral index changing from $1.67\pm 0.12$ to $2.24\pm 0.36$ 
(Blom et al.~1996). We interpret this complex spectral shape in terms of our 
simple one component jet model and Melia \& K\"onigl~(1989) geometrical disk 
model. It is assumed that the disk parameters has not been changed between 
these two observations.
The disk is defined by the inner disk radius $r_{in} = 10^{14}$ cm, 
the radius of the thick disk $r_{mid} = 2r_{in}$, its temperature 
$T_{mid} = 5\times 10^4$ K at $r_{mid}$, the exponent $\delta = 2$, 
and the opening angle of the thick disk $\beta =
80^o$. The fits to the spectra of PKS 0208-512 are shown in Fig.~2 during 
different observations (Blom et al.~1996). The spectra are obtained 
for the following parameters of the jet: $E_0 = 0.08$ V cm$^{-1}$ at the base 
of the jet $z_0 = r_{in}$, and $l_{\rm rec} = 0.005$. We assumed
that the injection rate of electrons is uniform along the jet. This might 
be motivated by the fact that the flux of the electrons in the jet should 
be conserved. 

The change in the spectra,
during two observations of PKS 0208-512, is caused only by changing 
the distribution of the electric field along the jet, defined by the parameter
$\alpha$ (see Eq.~3). Its value is equal to 1.4, for the observations during 
May - June 93, and
to 0.67, for the observations during July 91 - Jan 92. The change in these
spectra corresponds to different equilibrium Lorentz factors of electrons
during propagation in the electric field of the jet. 
In Fig.~3 we present the values of $\gamma_{\rm eq}$ as a function of the 
distance $z$ along the jet, for two cases 
discussed above. Note the difference in extraction of the electric field 
energy by electrons
propagating through the radiation fields of thick and thin disk in both cases.
This is the reason of anticorrelation of relative power emitted at the 
high energy (above $\sim 100$ MeV) and the low energy parts of the spectrum. 
In such model the cut-off in the spectrum, observed at around a few GeV 
(Kanbach~1996), is caused by inefficient
acceleration of electrons in the electric field of the particle dominated 
jet. This interpretation is different from another possible 
explanation of such a feature as being due to absorption
of high energy $\gamma$-rays in the soft radiation (e.g. Bednarek~1993, 
Sikora, Begelman \& Rees~1994, BK95, Blandford \& Levinson~1995, Pohl~1996).

\section{Conclusion}

We propose a simple geometric explanation as responsible for the formation of
the two component $\gamma$-ray spectrum in MeV blazars.
It is shown that in terms of the standard external comptonization model of the 
disk radiation, the two components in the $\gamma$-ray spectrum can be 
explained if the disk geometry changes from the geometrically thin one, in the 
outer part, to the geometrically thick one, in the inner part. Computed 
$\gamma$-ray spectra can well describe the complex shape of the most 
prominent MeV blazar
PKS0208-512 (Fig.~2). The model is able to explain the relative changes in the 
strength of the soft MeV component and hard component above $\sim 30$ MeV.
It is also consistent with the high energy cut-off in the $\gamma$-ray energy 
range. Since the lower $\gamma$-ray emission is produced closer to the central 
engine, it is likely that the GeV emission, produced further along the jet, is
delayed in respect to the MeV emission. This model can also simply explain the
appearance of a bump in the blazar spectra at energies below $\sim 0.511$ 
MeV (the case of 3 C273, McNaron-Brawn et al. 1997),
which can not be explained by the $e^{\pm}$ annihilation in the jet model. 
The peak in the $\gamma$-ray power spectrum, 
predicted by our model, is only determined by the balance between energy gains 
and energy losses of electrons propagating inside the
thick disk (see Fig.~1). Therefore, it may appear in principle, at arbitrary 
energy. 

It is assumed in this paper that the electrons are accelerated
rectilinearly as discussed originally by Bednarek \& Kirk~(BK95). However, 
we expect that similar two component spectra should appear if electrons,
accelerated by a relativistic shock propagating along the jet axis,
scatter radiation of the inner and outer disk. The acceleration of electrons 
can be also saturated by the ICS losses and the maximum energy of electrons is 
determined by the balance between acceleration efficiency and energy losses. 
Only the geometry of
the ICS scattering process by electrons distributed isotropically in the 
blob frame becomes more complicated as shown by e.g. Dermer \& 
Schlickeiser~(1993).

   \begin{figure}
      \vspace{7.2cm}
\includegraphics{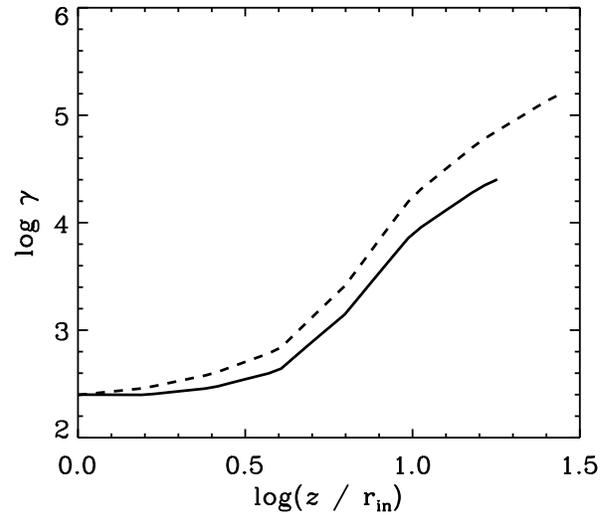}
      \caption[]{The equilibrium Lorentz factor of electrons as a function of 
the propagation distance along the jet for which the $\gamma$-ray spectra 
presented in Figs.~2 are obtained and compared with  May - June 93 observations (full curve) 
and July 91 - Jan 92 observations (dashed curve). }
\label{fig3}
    \end{figure}

\section*{Acknowledgements}
I thank Dr H. Blom for providing me with the data on the 
gamma-ray observations of PKS 0208-512 by the COMPTEL and EGRET 
detectors on Compton Gamma-Ray Observatory, the referee for comments, 
the editor for suggestion,
and Dr M. Ba\l uci\'nska-Church for reading the manuscript.

\end{document}